# Automated Small-Cell Deployment for Heterogeneous Cellular Networks

Weisi Guo, Siyi Wang[†], Xiaoli Chu[§], Jiming Chen[¶], Hui Song[¶], Jie Zhang[§]

University of Warwick, UK, Email: weisi.guo@warwick.ac.uk

[†] University of South Australia, Australia, Email: siyi.wang@mymail.unisa.edu.au

[§] University of Sheffield, UK, Email: {x.chu, jie.zhang}@sheffield.ac.uk

[¶] RANPLAN Wireless Network Design, UK, Email: {jiming.chen, hui.song}@ranplan.co.uk

## Abstract

Optimizing the cellular network's cell locations is one of the most fundamental problems of network design. The general objective is to provide the desired Quality-of-Service (QoS) with the minimum system cost. In order to meet a growing appetite for mobile data services, heterogeneous networks have been proposed as a cost- and energy-efficient method of improving local spectral efficiency. Whilst unarticulated cell deployments can lead to localized improvements, there is a significant risk posed to network-wide performance due to the additional interference.

The first part of the paper focuses on state-of-the-art modelling and radio-planning methods based on stochastic geometry and Monte-Carlo simulations, and the emerging automatic deployment prediction technique for low-power nodes (LPNs) in heterogeneous networks. The technique advises a LPN where it should be deployed, given certain knowledge of the network. The second part of the paper focuses on algorithms that utilize interference and physical environment knowledge to assist LPN deployment. The proposed techniques can not only improve network performance, but also reduce radio-planning complexity, capital expenditure, and energy consumption of the cellular network. The theoretical work is supported by numerical results from system-level simulations that employ real cellular network data and physical environments.

## I. INTRODUCTION

Traditionally, cellular network deployment has been primarily designed for outdoor coverage and voice services, which are achieved by overcoming the stochastic nature of the radio propagation environment. In the past decade, there has been an unprecedented growth in mobile data demand. This has led to revolutions in the multiple-access technology, as well as an increase in cell density and spectrum reuse. The 3rd and 4th Generation cellular networks mostly employ full bandwidth reuse (reuse pattern one), and the cell density in urban areas is in excess of 6 cells per square kilometer per operator. This has yielded a system-level capacity that is largely interference-limited, as opposed to propagation-limited.

Mobile data demands in the cellular networks occur predominantly (70%) in indoor areas, while the traditional radio-planning strategy is ill-equipped to address this issue. The indoor coverage issue is especially challenging for large buildings such as shopping malls, hotels, enterprise and government offices, where multiple indoor surfaces of different electromagnetic properties impede signal propagation. The typical indoor subscriber density in the aforementioned buildings is high, but the quality-of-service (QoS) delivered to them is currently low.

Three factors motivate cell planning optimization: interference, user location, and radio propagation. Whilst a lot of work has gone into signal processing and resource management techniques for mitigating interference, there have been less efforts on the latter two issues. In this paper, we investigate how to optimize the cell location subject to the interference pattern, given a certain user distribution and radio propagation model.

Low-power nodes (LPNs), such as femto access points (FAPs) and relay nodes (RNs), have been proposed as low-cost and low-energy methods for improving local spectral efficiency [1]. Such LPNs are integrated into the existing cellular network via wired broadband (e.g., ADSL, optical-fibre), or wireless backhaul (e.g., in-band transmission, or microwave links). The resulting cellular network is known as a heterogeneous network (HetNet).

A key challenge to such a HetNet is how to mitigate the excessive interference in areas that traditionally would have good coverage, but now suffer degradation due to the additional inference created by nearby LPNs [2] [3]. In Fig. 1, the mean received signal-to-interference-plus-noise ratio (SINR) in an example HetNet is shown. The HetNet consists of a sectorized macro base station (BS) with 12 LPNs deployed within its coverage area. It can be seen that the SINR is high near the LPNs, but rapidly falls to a level that is below the original macro-BS serving SINR in regions surrounding the LPN coverage areas. This is due to the excessive cross-tier interference.

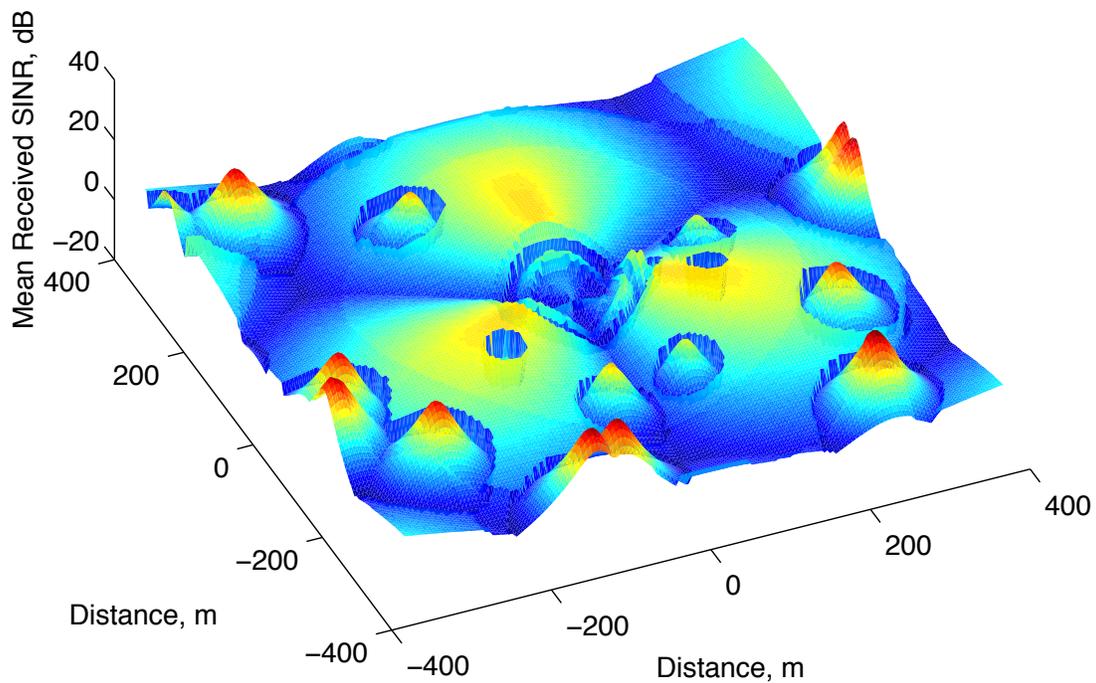

Fig. 1. HetNet with a macro-BS and randomly deployed Femto-cells. Femto-cells improve local signal strength but severely degrade the surrounding-area signal strength due to excessive interference.

## II. Heterogeneous Network Planning

### A. Modelling

In order to optimize the cellular network performance through cell-site planning and transmission techniques, different modeling approaches have been taken over the years to characterize the network performance:

- Monte-Carlo Multi-Cell Model (Simulation): can include multiple effects, which are not easily describable by tractable mathematical functions, such as ray-traced pathloss models, antenna patterns, terrain, clutter, and cell specific configuration data. For specific models, a large volume of data is required and an example is shown in Fig. 2a), where the mean received downlink signal power is from a major operator's HetNet in a European city, with 95 macro- and pico-BSs modeled. For generic models, a hexagonal cell layout with wrap-around is typically employed to obtain an upper-bound of network performance [4].
- Stochastic Geometry Model (Statistical): can capture the network-wide performance of a non-uniform network deployment [5], but includes only stochastic effects that are mathematically tractable. An example for a network with a certain cell density is shown in Fig. 2b).
- Single Cell Linear Model (Deterministic): can capture the specific performance variations across the coverage area of a single cell in a multi-cell network [6]. An example is shown in Fig. 2c), where the a framework considers only a dominant interference source, which clearly has limitations. Provided that there is always a dominant interference source, scalability in cell density is not an issue. If scaling the network means that more and more locations suffer equal interference from multiple sources, then the linearity of the model will break down.

Fig. 3 plots the cumulative distribution function (CDF) of the received downlink SINR across a network with the three modeling approaches. Simulation parameters for the realistic network (in a European city) are: 96 realistic Macro- and Pico-BSs in a 9 km × 6 km area, with ray-traced pathloss models (PACE 3G) and realistic antenna patterns. Parameters for the theoretical models employ the WINNER Urban statistical pathloss model and omni-directional antenna patterns. This work was conducted at the University of Sheffield with the Mobile VCE (MVCE) and multiple industrial partners [7].

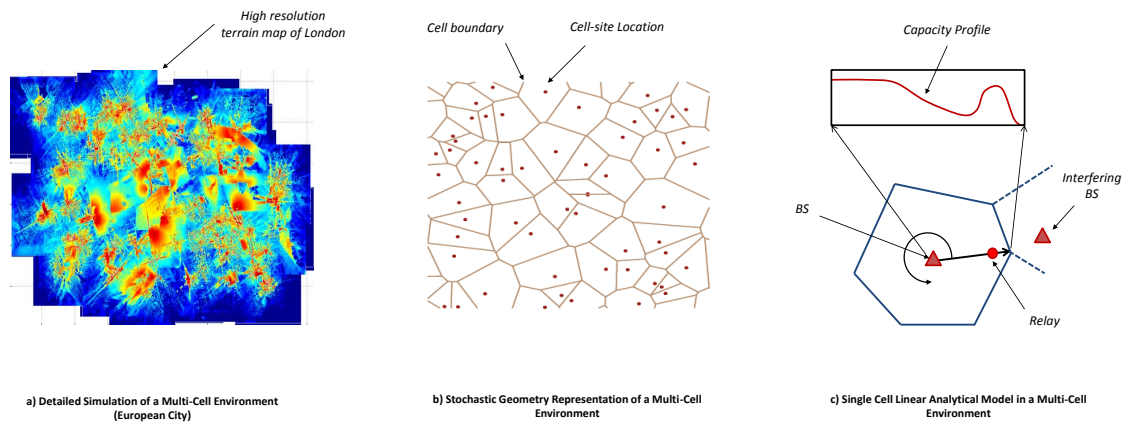

Fig. 2. Heterogeneous network modeling methods: a) Monte-Carlo simulation of a realistic environment; b) stochastic geometry representation of a multi-cell network; c) linear model of a single cell in a multi-cell network.

We can see that if the realistic European city's network is taken as a reference, then the stochastic geometry is quite accurate. The hexagonal and linear models can use a back-off factor to improve their accuracy. The relative merits of each modeling technique are beneficial for different purposes. Specific challenges typically warrant the use of simulation based approaches, where custom features can be accommodated. Stochastic models can yield insights on the impacts of cell density, transmit power and pathloss, but they are not well suited to analyze effects that are not easily modeled by probability distributions such as vertical antenna patterns and terrain clutter. Furthermore, stochastic models only provide a statistical deployment solution (e.g., the optimal average number of femtocells per macrocell), as opposed to a deterministic deployment solution (e.g., the optimal number and locations of femtocells in a *specific* macrocell). The linear model offers a balance between the aforementioned two approaches by providing a deterministic deployment solution in a way faster than simulations.

## B. Cell-site Planning and Challenges

Cell-site planning has traditionally targeted coverage percentage and traffic density. The latter is difficult to characterize, especially given its dynamic nature and the shifting trends in usage patterns and social mobility. Nonetheless, a great deal of traffic information is inferred and forecasted from:

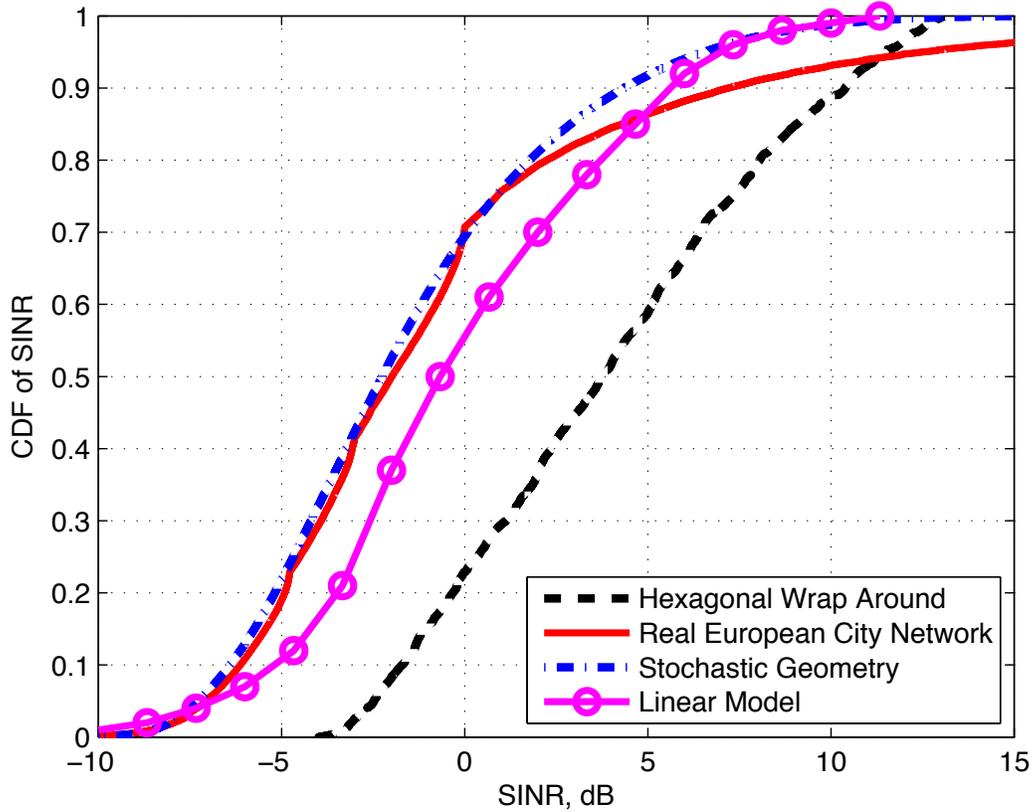

Fig. 3. CDF of the network-wide SINR with different modeling approaches.

- Demographic Data: the residential and business population distribution based on demographic census;
- Traffic Data: vehicular data based on public transport and private vehicle movement patterns;
- Fixed Line Data: based on correlation with fixed line telephony records, given that most mobile data traffic occurs indoors.

On a macro- and statistical-scale, the stochastic framework introduced in [5] can calculate the LPN density as a function of the transmit powers, statistical pathloss exponent, and noise level.

For radio planning on a micro scale, Monte-Carlo simulations are employed along with detailed urban terrain maps and ray-traced pathloss models. This is recognized as an NP-hard problem. Given a set of possible cell-site or LPN locations, iterative techniques are usually used to scan the optimal locations for cell-sites and LPNs. Optimization methods such as integer programming, simulated annealing, and multi-era genetic programming algorithms are employed to search for optimal solutions. Meta-heuristic methods such as Tabu search [8] can accelerate the process by ignoring previous negative search results (within a certain iteration period) that are stored in a memory. The ultimate deliverable goal is to make the search complexity linearly proportional to the number of BSs and user equipments (UEs) considered.

To give an idea of the scale and complexity of the challenge, a typical developed urban metropolis has approximately 2 BS sites per square kilometer per operator. This equates to approximately 100 BSs per city, incorporating over 300 macrocells. In order to deploy LPNs in a HetNet, investigations carried out by the industry have shown that the typical number of LPNs required to boost indoor coverage to outdoor levels, ranges from 30 to 100 per BS, yielding a lower-bound of 60 cells per square kilometer and 3000 cells per operator in a city.

The resulting radio planning complexity for the HetNet is extremely high, primarily because:

- *Cell Densification*: 30 to 100 fold increase in cells;
- *Coverage Resolution*: 100 fold increase (from 20m to 2m) in coverage resolution for LPNs and indoor areas, and at least a 3 fold increase in coverage height resolution;
- *Indoor-Outdoor Pathloss Complexity*: unknown increase in computation time;

which lead to at least a 10000 fold increase in the computation time for radio coverage analysis or prediction. This would increase deployment planning and more importantly system optimization times to unfeasible levels. There is therefore a temptation to deploy LPNs without articulated radio planning and rely on signal processing techniques to improve performance. The danger with this approach is that in the absence of effective interference mitigation techniques, there might be zones of intense interference as shown in Fig. 1.

The complexity of deploying LPNs and predicting their performance can be reduced by finding approximate deployment locations using key network parameters. In order to avoid or reduce the complexity of protracted simulations, analytical methods such as the stochastic geometry model proposed in [5] can be used. Whilst stochastic geometry offers offer network-wide mean performance bounds that relate to node density and other parameters, the challenge of how to plan each specific BS of a HetNet remains open.

*C. Towards Automated Deployment*

In order to gain insight of LPN deployment location on a single BS level, the latest development in network performance modelling has included the effects of:

- *Interference*: from the co-channel transmission of a dominant neighboring cell [6];
- *Capacity Saturation*: realistic transmission schemes suffer from mutual information saturation in discrete modulation schemes. For example, in the LTE physical layer, the maximum achievable spectral efficiency is 4.3 bits/s/Hz for a typical outdoor environment. Research in [9] has shown that existing solutions, which do not consider spectral efficiency saturation, lead to a significant waste in radio resources.

The work in [6] shows that by jointly considering the effects of interference and capacity saturation, the optimization solution is significantly different from those of noise-limited channels without capacity saturation [10]. Automated cell deployment is a concept that attempts to deterministically find the optimal location of a new cell, subject to knowledge about the locations of existing cells, users and the propagation environment. This is in contrast to random deployment or optimization using brute-force search methods in simulations. Whilst some of the automated deployment solutions are known to experienced radio-planning engineers, the availability of the deployment location in closed-form as a function of transmit power, transmission scheme and pathloss parameters, is novel and significantly beneficial. The work has been applied to: outdoor wireless relays [6], and access-points (APs) for indoor areas [11], [12]. The automated deployment model has been validated against an industrially bench-marked multi-cell system simulator. The following sections provide an overview of the automated deployment model and its impact on the future of HetNet planning.

## III. Automated Outdoor Deployment

*A. Motivation and Methodology*

For the outdoor cellular network, one of the largest sources of operational expenditure is the tethered back-haul rental cost. Furthermore, the dense nature of outdoor LPNs requires the operator to balance the optimal-coverage LPN locations with the availability of back-haul cabling. These have motivated the deployment of wireless RNs. However, the challenge with allocating scarce spectrum to relaying is not only difficult to manage, but also complex to optimize.

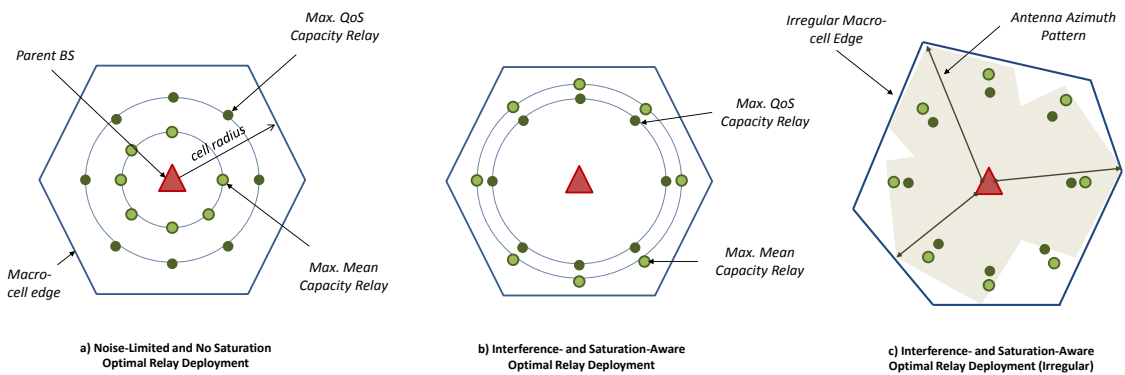

Fig. 4. Optimal RN deployment for: a) noise-limited and saturation-free channels; b) interference-limited and saturation-aware channels; c) interference-limited and saturation-aware channels with irregular cell coverage.

For automated cell deployment, the optimal location of a cell (Femto-cell or Relay-Node) is deterministically found using an algorithm. A linear model proposed in [6], [12] uses the estimated signal power received from each cell. The estimation process considers the transmit power, statistical pathloss, and cell location. The effects of terrain clutter and antenna patterns have not yet been considered. However, a realistic system can also measure the real signal power received from different BSs. The measurements can be used instead of the estimation method. The measurements can then be used to optimize the locations of cells in accordance with the formulas devised in [6], [12].

*B. Theory*

In [6], the proposed theoretical framework for wireless RNs accounts for the effects of interference and capacity saturation. The optimal locations of RNs from their parent BS are fundamentally different to those in a Gaussian noise channel [10]. As shown in Fig. 4a), in a noise-limited and saturation-free channel, the optimal parent-BS to RN distance is to deploy the RNs relatively close to the parent BS, so that the BS-RN channel could be good enough to not limit the RN-UE channel. However, this may create two problems in a realistic network:

- UEs that are close to the BS already experience close to saturated performance and do not require relaying;
- RNs are likely to degrade that saturated performance through in-band interference, whilst offering very little improvement.

In fact, it was found that unarticulated or mis-calculated deployment of LPNs may cause a network-wide spectral efficiency degradation.

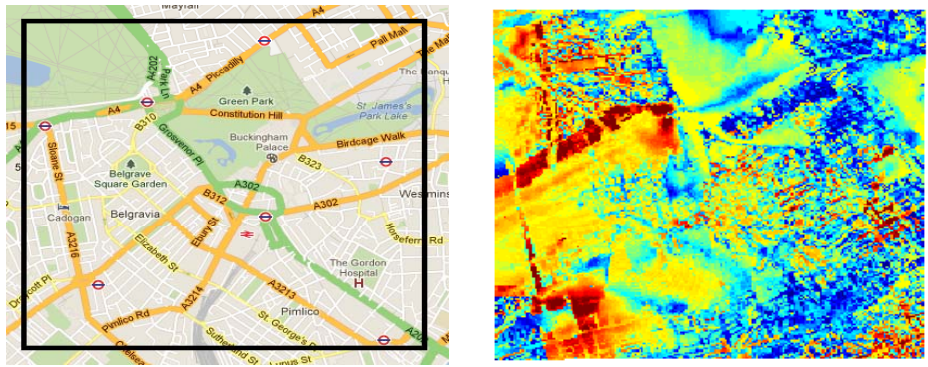

**a) Central Area of a European City: HetNet Capacity Map**

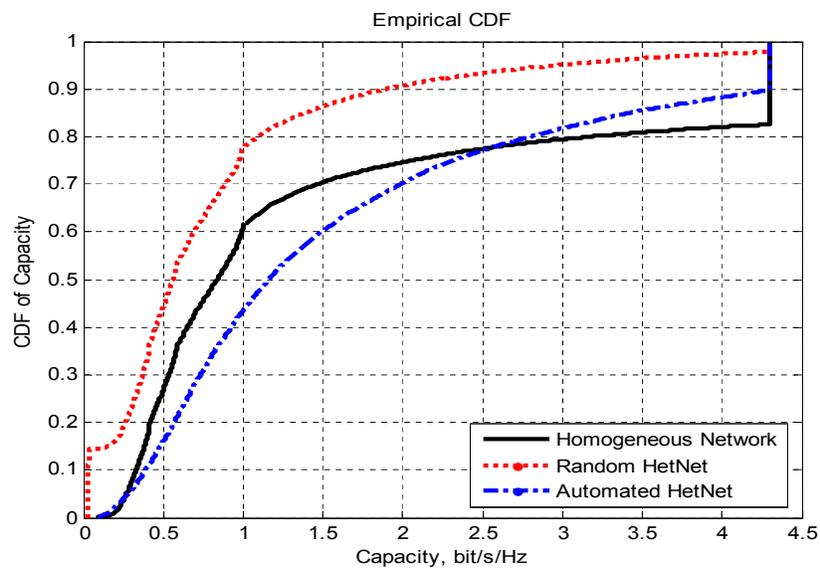

**b) CDF of Capacity Across City Area**

Fig. 5. European city's spectral efficiency profile for a HetNet: a) location and spectral efficiency map; b) CDF of spectral efficiency profile [7].

GUO *et al.*: AUTOMATED SMALL-CELL DEPLOYMENT FOR HETEROGENEOUS CELLULAR NETWORKS 11As shown in Fig. 4b), in an interference-limited and saturation-aware channel, the optimal RN location is approximately 0.7-0.8 of the macrocell coverage radius ($d_{\text{BS}}$) away from the BS. The optimal BS-RN distance ($d^*_{\text{BS-RN}}$) that maximizes the mean network spectral efficiency can be expressed as [6]:

$$d^*_{\text{BS-RN}} \propto d_{\text{BS}} \left( \gamma_{\text{Sat.}} \frac{P_{\text{RN}}}{P_{\text{BS}}} \right)^{-\frac{1}{\alpha}}, \tag{1}$$

where $\gamma_{\text{Sat.}}$ is the signal-to-noise ratio (SNR) at which the capacity saturates, $\alpha$ is the pathloss distance exponent, and $P_{\text{RN}}$ and $P_{\text{BS}}$ are the transmit power of the RN and the BS, respectively. The expression shows that the optimal distance from the RN to the BS is inversely proportional to the transmit power ratio of the RN and the BS, and the constant of that proportionality is the pathloss distance exponent. The optimal BS-RN distance is also transmission scheme aware, since a lower-order transmission scheme such as binary phase shift keying (BPSK) (with lower $\gamma_{\text{Sat.}}$ value) leads to the RNs being deployed further away from the BS, in order to protect UEs that already experience saturated performance. The proposed RN deployment yields an optimal balance between improving the BS-RN channels and improving the performance of cell-edge UEs.

Another parameter of concern is the number of RNs per BS sector that maximizes the spectral efficiency of the network. The interference- and saturation-aware theoretical framework in [6] shows that the optimal number of RNs per BS sector ($N^*_{\text{RN}}$) is upper bounded by

$$N^*_{\text{RN}} \leq \pi \left( 2 \frac{P_{\text{RN}}}{P_{\text{BS}}} \right)^{-\frac{1}{\alpha}}, \tag{2}$$

which shows that the optimal number of RNs per BS sector is inversely proportional to the transmit power ratio of the RNs to the BS, and the constant of that proportionality is the pathloss distance exponent.

Furthermore, due to the radial nature of the RN deployment framework, the result can be extended to non-uniform cell geometries with azimuth antenna patterns, as shown in Fig. 4c). The network-wide spectral efficiency improvement achieved by the proposed automated RN deployment over the random deployment is approximately 55% for outdoor RNs [6]. Whilst the automated RN deployment solutions are known to experienced radio-planning engineers, the availability of the solution in closed-form as a function of transmit power, transmission scheme and pathloss parameters, is novel and of benefit by reducing radio-planning time.

*C. Validation Using Real Network Data*

In order to validate the automated LPN deployment solutions in a realistic outdoor cellular network, the automated deployment algorithm is applied to data from a real cellular operator's network in a developed urban city [7]. Fig. 5a) shows the area of focus (data extraction), which is a 4 square kilometer area in central urban area, including approximately 4 macro-BSs and 40 LPNs. The interference from 92 other BSs in the city area is also considered.

The results in Fig. 5b) show that the unarticulated random LPN deployment actually degrades the network performance as compared to the homogeneous deployment, which was also predicted in [6]. Articulated auto-deployment of LPNs on the other hand achieves a significant improvement in network-wide spectral efficiency, against both the random LPN deployment and the conventional homogeneous cellular network. In terms of mean spectral efficiency, the improvement is approximately 50%, which closely matches the theoretical predictions found in [6].

## IV. Automated Indoor Deployment

*A. Motivation*

In indoor areas, the availability of tethered backhaul makes the deployment of LPNs or APs an attractive solution. Whilst the locations of outdoor cells are controlled by operators to meet network performance targets, there is less understanding or control on where indoor LPNs should be placed. Conventionally, indoor APs are deployed at locations of convenience. Indeed, the end-user can not always arbitrarily decide where an AP can be placed. Recent research shows that in a strong interference environment, some regions of a room are more beneficial than others. The optimal placement of APs has been previously investigated in [13], whereby iterative computation techniques were used to find the optimal locations of multiple nodes in an indoor environment [11]. More recently, there has been development on how to use statistical pathloss and user distribution parameters to predict the optimal location of an AP [12], without using exhaustive computational algorithms.

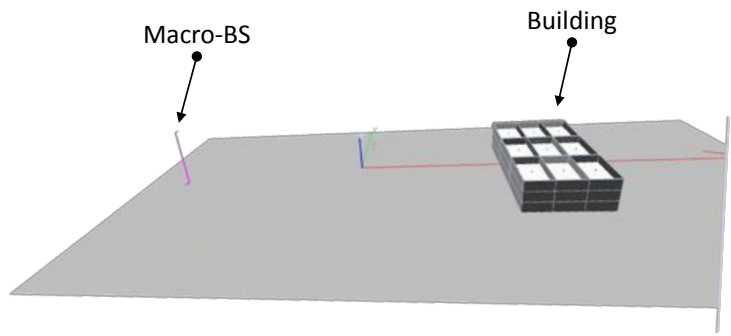

**a) Network Setup in iBuildNet Simulation Tool**

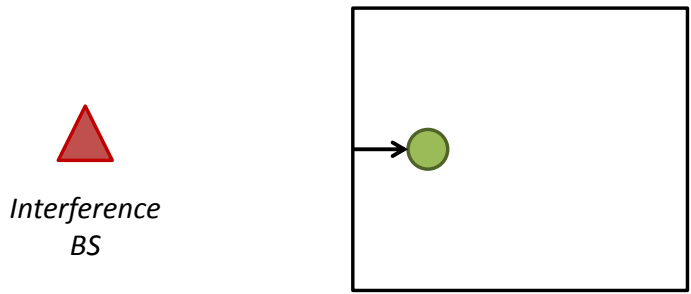

**b) Single-Room: Single FAP Optimal Placement**

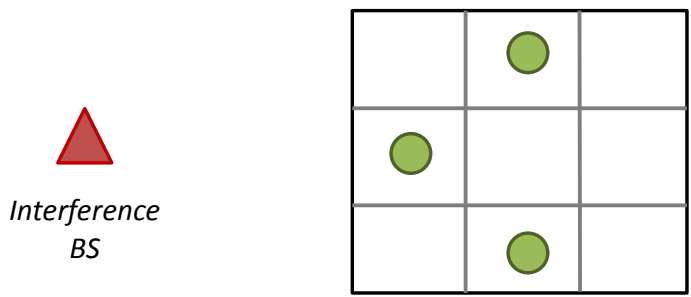

**c) Multi-Room: Multi FAP Optimal Placement**

Fig. 6. Optimal FAP deployment for maximum uniform coverage in: a) investigation setup in iBuildNet; b) a single room; c) multiple rooms.

*B. Theory and Validation*

Recent work in [12] considered both 802.11n WiFi APs and LTE FAPs, for single- and multi-room buildings. The key findings are, for maximizing mean spectral efficiency, the APs should be deployed with the following steps (using FAPs for example):

1) A single FAP should be deployed adjacent to the external wall that faces the closest outdoor macro-BS, as shown in Fig. 6a) and Fig. 6b), and on the building floor that most closely matches the height of the outdoor macro-BS [11]. The location knowledge of the macro-BSs can be found through government cell databases. Based on spectral efficiency maximization, the optimal location ($d^*_{\text{FAP}}$) can be explicitly found [12]:

$$d^*_{\text{FAP}} \approx d_{\text{building}} \left[ 1 + W^{-\frac{2}{\alpha}} (1 + \frac{d_{\text{building}}}{d_{\text{FAP-BS}}}) \right]^{-1}, \quad (3)$$

where the optimal location $d^*_{\text{FAP}}$ is taken as the distance from the external wall nearest to the outdoor macro-BS, $W$ is the aggregate penetration loss of the internal walls, $d_{\text{building}}$ is the length of the building, and $d_{\text{FAP-BS}}$ is the distance between the serving FAP and the nearest dominant interfering BS.

2) If more than one FAP is deployed, one FAP should be deployed as described above, the other FAPs should be placed at maximum mutual distance, so that interference between FAPs is minimized [11], as shown in Fig. 6c).

The optimal number and locations of FAPs should be determined sequentially, from the lowest number to the highest. In a building with a small number of rooms, inter-FAP interference dominates the indoor network performance [11].

The above theoretical automated deployment algorithm employs statistical pathloss expressions. The results have been validated against an outdoor-indoor system simulator known as iBuildNet [14], using ray-traced pathloss models both outdoors and indoors. The simulation configuration is shown in Fig. 6a), and the results found strongly agree with those predicted by the theoretical automated deployment algorithm. The theory therefore allows two potential benefits: automated deployment of cells that are conscious of mutual interference, and providing an initial location input for more protracted simulation-based deployment optimization software.

## V. Energy and Cost Savings

A key benefit of deploying a more spectrally efficient network is so that the carbon footprint and expenditures are reduced. There is already a significant commitment from major wireless operators to cut their carbon footprint and reduce operational expenditures.

Whilst the total power consumption of a LPN is typically small (10-25W), there are already hundreds of millions of LPNs across the world, and this figure is set to grow rapidly. Therefore it is important to consider their ecological and economical impact. Using bench-marked system simulation tools, it was found that the network-wide spectral- and transmit energy-efficiency improvement achieved by the proposed automated deployment over the random deployment is approximately 20-50%, depending on the environment [6], [11], [12]. This leads to a carbon footprint reduction of 7-16% and a small operational expenditure (OPEX) saving of 5-12% [7].

Furthermore, as a result of deploying LPNs more efficiently, it can be argued that fewer LPNs need to be deployed to achieve the same mean network performance than the reference system (random deployment). In that case, both the energy and cost savings are more profound and can reach 40-50% [4] [7].

## VI. Future Work

To the best of our knowledge, relevant cell self-deployment and self-organization work has been conducted mainly by Bell-Labs and other European researchers for BSs that can fly or at least reposition themselves in some way [15]. However, it is not yet clear from their work how and where the cells will reposition themselves and how the mutual optimisation works. The work conducted in self-deployment provides that insight. Coupled with certain automated mechanisms, in the future cells can reposition themselves in accordance to user patterns, traffic loads, and interference conditions.

One of the key challenges with deployment optimization generally is that the optimal capacity location of a cell, may not be available for practical and economic reasons. In that case, each node should be equipped with certain self-optimization features such that sub-optimal placement does not exacerbate the network performance. Another reality is that there is a complex balancing act between profit margins from capacity improvements and those from savings made to site rental costs.

## VII. Conclusions

This paper has given a survey and tutorial of emerging work on deploying LPNs in a HetNet. Due to the high node density of future HetNets, there is a demand for solutions that can reduce radio network planning and potentially allow both outdoor and indoor nodes to be deployed autonomously or with very little guidance. Recent advances in interference- and saturation-aware deployment algorithms can potentially enable LPNs to be deployed whilst minimizing inter-cell interference and maximizing network-wide spectral efficiency. The theoretical work in this area is validated with simulation results employing realistic network and environmental data.

The results show that deploying LPNs without location optimization can degrade network-wide spectral efficiency, while automated deployment optimization techniques can provide a low complexity solution to intelligent HetNet roll out.

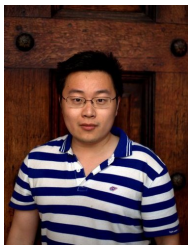

**Weisi Guo** received his M.Eng., M.A. and Ph.D. degrees from the University of Cambridge. He is currently an Assistant Professor and Co-Director of Cities research theme at the School of Engineering, University of Warwick. He is the author of the VCEsim LTE System Simulator, and his research interests are in the areas of heterogeneous networks, self-organization, energy-efficiency, and cooperative communications.

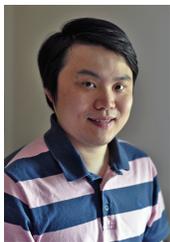

**Siyi Wang** received the B.Eng. in Communication Engineering from Shanghai University in 2006 and M.Sc (with distinction) degree in Electrical Engineering from the University of Leeds in 2007. He is currently studying a Ph.D in Institute for Telecommunications Research at the University of South Australia. The research for his Ph.D investigates communicating data via the diffusion of nano-particles and is funded by University President's Scholarships. His research interests include: indoor-outdoor network interaction, small cell deployment, machine learning, stochastic geometry, theoretical frameworks for complex networks and nano-based particle communications.

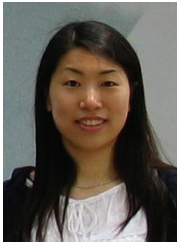

**Xiaoli Chu** is with the EEE Department at the University of Sheffield. She received the PhD degree from the Hong Kong University of Science and Technology in 2005. From 2005 to 2012, she was with the Centre for Telecommunications Research at Kings College London. She has published more than 50 journal and conference papers. She is the leading editor of the book Heterogeneous Cellular Networks (Cambridge University Press, May 2013), and guest editor of the Special Issue on Cooperative Femtocell Networks for ACM/Springer Journal of Mobile Networks & Applications in 2012.

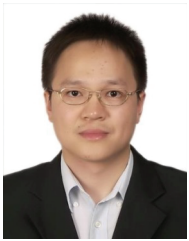

**Jiming Chen** received his M.S. and Ph.D degree in communication and information system from University of Electronic and Science Technology, China in 2003 and 2006, respectively. From 2006 to 2011, he was a Research Scientist and Member of Alcatel-Lucent Technical Academy in Bell labs at Alcatel-Lucent Shanghai Bell Ltd.. Currently, he is a Senior Research Fellow at Ranplan Wireless Network Design Ltd. (www.ranplan.co.uk) that produces the world leading HetNet planning and optimisation tools - iBuildNet.

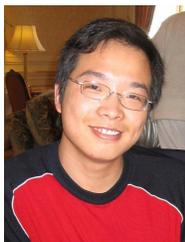

**Hui Song** is currently the R&D Manager at Ranplan Wireless Network Design Ltd. (www.ranplan.co.uk), UK. He contributed significantly to Ranplan's in-building wireless network design and optimisation tool iBuildNet.

He obtained his Ph.D in Wireless Communications from University of Bedfordshire in April 2010. His research interests are on the fields of: wireless network planning and optimization techniques, next generation wireless systems, HetNet and small cells (femtocells, picocells and metrocells etc.), MIMO-OFDM link adaptation, propagation modelling and system-level simulation.

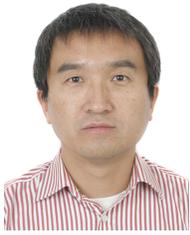

**Jie Zhang** is a full professor and holds the Chair in Wireless Systems at the Department of Electronic and Electrical Engineering, University of Sheffield, UK. His research interests are focused on radio propagation, indoor-outdoor HetNet planning and optimisation, small/femto cell, self-organising network (SON) and smart environments. Since 2006, he has been an Investigator of over 20 research projects worth over 17 million GBP (his share is over £5.0 million) by the Engineering and Physical Science Research Council (EPSRC), the European Commission (EC) FP6/FP7 and the industry etc. He was/is one of the Investigators of some of the earliest projects on Femtocell, wireless friendly building and green communications.

Since 2007, he has published over 100 papers in referred journals and conferences (e.g., IEEE Trans. on WC/Communications/Antenna and Propagation/Microwave Theory and Techniques, IEEE JSAC/Comm. Mag.). He is a lead author of the books "Femtocells: Technologies and Deploymen" (Wiley, Jan. 2010) and"Small Cells: Technologies and Deploymen" (Wiley, Q2 2013). He and his colleagues published one of the most widely cited femtocell papers"OFDMA femtocells: A roadmap on interference avoidanc" and some early work on Femtocell self-organisation.